# A NUMERICAL ANALYSIS OF THE MODIFIED KIRK'S FORMULA AND APPLICATIONS TO SPREAD OPTION PRICING APPROXIMATIONS

**AUTHORS:**

SUREN HARUTYUNYAN

ADRIÀ MASIP BORRÀS

**PROJECT'S ADVISOR:**

ELISA ALÒS ALCALDE

*Abstract*

*In this paper we study recent developments in the approximation of the spread option pricing. As the Kirk's Approximation is extremely flawed in the cases when the correlation is very high, we explore a recent development that allows approximating with simplicity and accuracy the option price. To assess the goodness of fit of the new method, we increase dramatically the number of simulations and scenarios to test the new method and compare it with the original Kirk's formula. The simulations confirmed that the Modified Kirk's Approximation method is extremely accurate, improving Kirk's approach for two-asset spread options.*

*Keywords:* Spread options; Spark spread options; Kirk's formula, Modified Kirk's formula.

# INDEX

# I. Introduction.

With the increasing sophistication of the financial markets and the surging popularity of the energy markets, innovative methods are needed, to be able to understand the inherent risks of new financial products, and to be better positioned to manage new uncertainties. This paper is devoted to the study of the spread options in the context of the energy markets, by studying the modification of the classical Kirk's Approximation, presented by Alòs and León[1]. While the Modification of the Kirk's Approximation, presented a simple, but a very accurate method to approximate option prices, the original paper, introduced few examples and simulations to analytically study its goodness of fit. For this reason, the authors want to expand the examples and trials of the original papers, to assess the behavior of the Modification of the Kirk's Approximation in a wide range of cases and situations.

The organization of the paper is as follows. In section 2, we will expose the regulatory framework of the energy markets, in particular, the electricity market in the European Union, starting with the evolution of the regulation. Section 3 introduces the determinants of the electricity price volatility by analyzing the demand and the supply in the case of the electricity markets. Section 4 is devoted to derivatives, particularly options, as financial instruments which allow us to hedge the inherent volatility of the electricity price. Section 5 explains the latest development of the further deregulation of the energy markets, and to be precise, the electricity markets, by the European Commission.

In section 6, we introduce the general pay-off form of options with random strikes, such as two-asset spread options. We continue in section 7, with a brief review of the Margrabe's formula, which we will see later in the section that it is a special case of the Kirk's Approximation, discussed there. We add in section 8 the framework and the definitions of the Modified Kirk's Approximation. Finally, in section 9, we introduce the Monte Carlo simulations, in order to be able to compare these to the results obtained by the Kirk's Approximation and the Modified Kirk's Approximation. Here we introduce more trials, and scenarios, and we compare the errors of the Kirk's formula to the Modified Kirk's formula to assess the goodness of fit of the latter. An Annex with the key parts of the code is added after the delivery of the conclusions.

---

[1] Alòs, E., & León, J. A. (2015). On the short-maturity behaviour of the implied volatility skew for random strike options and applications to option pricing approximation. *Quantitative Finance, 16*(1), 31-42. doi:10.1080/14697688.2015.1013499. Retrieved from http://dx.doi.org/10.1080/14697688.2015.1013499.

## II. Regulation of the energy markets (focus on the electricity market)

Historically, the energy market and, in particular, the electricity market, have been under the control of the State. However, in the 1980s a wave of deregulation emerged. Specifically, in the early 1990s, the European Union (onwards, EU) started to pass directives in order to deregulate or liberalise the electricity market, i.e., open it to competition. This would generate an efficient and integrated energy market that, in turn, would bring about the so desired single market. Moreover, the more competition is introduced within the EU, the more this market will be able to compete with other markets (and so the EU's economy will be enhanced. Thus, complying with the Lisbon Strategy's 2000 aim)[2].

The purpose of the Single European Act of 1987 to reach a single market carried with it the need to deregulate the energy market. However, it was with the publication of the 1995 Green Paper on energy policy when the European Commission clearly took a stand regarding the energy market; it went for deregulation. Thus, in the next years (1996, 2003 and 2009) three European directives were passed in order to liberalise the electricity market.

In this sense, opening the electricity market to competition does not only imply moving away from public ownership of electricity assets towards more private involvement. It also requires unbundling generation and supply activities from transmission (where the power is transferred over a high-voltage net) and distribution (where the power is transferred over a low-voltage net) services. That is, real competition ought to be promoted in all sectors of the physical electric network. For instance, if generating and retailing companies are not on equal footing to have access to the network, then they won't have the same opportunities to place their product in the market. Therefore, equal and non-discriminatory access to the network should be guaranteed.

To this end, the European directives of 1996 and 2003[3] promoted legal and functional unbundling of transmission and distribution networks from activities of generation and supply. Moreover, independent, impartial and transparent regulatory agencies were created at the national level so as to ensure that the tariffs imposed by transmission and distribution operators to third parties – in order to get access to the network – are non-discriminatory. These tariffs ought to be cost-reflective, i.e., they shall allow to recoup the costs of operation, to maintain the network and to carry out the necessary investments in order to assure the viability of the network[4]. In this regard, the reliability of the network is crucial, so as to ensure that the power is transferred to where it's needed and to fully exploit the generation and supply capabilities[5]. Furthermore, the national regulatory authorities enjoy a margin of discretion when assessing costs incurred and claimed by transmission and distribution operators. In sum, these authorities can consider that some of the costs claimed could be avoided, and so the tariffs shouldn't reflect them. Thus, they boost network operators to keep avoidable costs down and to be more efficient.

---

[2] Dorsman, A. B. (2011). *Financial aspects in energy: A European perspective*. Heidelberg: Springer; pp. 11-14.

[3] Directive 96/92/EC of the European Parliament and of the Council of 19 December 1996
[3] Directive 96/92/EC of the European Parliament and of the Council of 19 December 1996 concerning common rules for the internal market in electricity and Directive 2003/54/EC of the European Parliament and of the Council of 26 June 2003 concerning common rules for the internal market in electricity and repealing the former.

Later, the European directive of 2009[6] introduced important changes as regards with unbundling of transmission and distribution networks from activities of generation and supply and, on the other hand, with reference to national regulatory agencies. In connection with the former, it established ownership unbundling of transmission system operators. This implies the appointment of the network owner as the system operator and its independence from any supply and production interests. Moreover, the same person should not be a member of the managing boards of both a transmission system operator or a transmission system and an undertaking performing any of the functions of generation and supply. Nevertheless, the directive maintained legal and functional unbundling of distribution system operators, as the scope for discrimination as regards third-party access is less significant.

Secondly, and turning to the national regulatory agencies, the directive's purpose was to harmonise their powers throughout the member states and to strengthen their independence from government. Furthermore, it should be noted that these public agencies, thanks to the new legislation, would be empowered to refuse certification to transmission system operators when network activities and generation and supply activities do not remain independent from each other, i.e., when these operators do not comply with unbundling rules.

---

[4] Energy Community. (2012). Regulated Energy Prices in the Energy Community – State of Play and Recommendations for Reform; p. 4. Retrieved from https://www.energy-community.org/portal/page/portal/ENC_HOME/DOCS/1568177/0633975AB98C7B9CE053C92FA8C06338.PDF

[5] Mack, I. M. (2014). *Energy trading and risk management: A practical approach to hedging, trading and portfolio diversification*. Hoboken: John Wiley & Sons; p. 10.

[6] Directive 2009/72/EC of the European Parliament and of the Council of 13 July 2009 concerning common rules for the internal market in electricity and repealing Directive 2003/54/EC.

However, despite the European directives' aim of achieving a single energy competitive market, currently the EU electricity market is, in part, still regulated. Specifically, in most of the EU Member States, the electricity generation market (wholesale market) is open to competition, whereas the electricity supply market (retail market) is subject to regulation. Indeed, if we compare it with the US regulation system of the electricity sector, the same picture is displayed. That is, the Federal government has deregulated the wholesale electricity market, but most of the States still fix the retail sales[7].

On the other hand, it should be noted that it's not economically feasible to build competing networks to serve the same customers. That's the reason why the networks transporting electricity (transmission and distribution) are considered natural monopolies. Therefore, the goal of creating a competitive electricity market ought to be restrained to the services activities (generation and supply)[8].

At present, Member States regulate the supply electricity market on basis of article's three provision of the European directive of 2009. This clause allows Member States to impose public service obligations on undertakings operating in the electricity sector. These obligations may relate, among others, to security of supply and to regularity, quality and price of supplies. Moreover, Member States will use this prerogative (which constitutes an exception to free market) as long as end-users' right to get access to electricity of a specified quality at reasonable prices is in danger.

In this sense, opening the retail market to competition would expose end-users to price volatility. Accordingly, most of the Member States dislike it, since it could deprive some consumers of getting access to electricity. For example, think of periods of peak demand, when the increase in electricity prices could prevent some end-users of affording this commodity. Importantly, this perception is even stronger in countries where, traditionally, electric utilities have belonged to the State for many years (public ownership). Therefore, most of the Member States prefer to regulate retail prices. In fact, this can explain why the degree of implementation of the European directives (and so the degree of liberalization of the electricity market) varies from country to country.

---

[7] Pillsbury Winthrop Shaw Pittman LLP. (2010). Electricity, Oil and Gas Regulation in the United States; p. 188. Retrieved from http://www.ourenergypolicy.org/wp-content/uploads/2013/08/ElectricityOilandGasRegulationintheUnitedStates.pdf

[8] Energy Community. (2012). Regulated Energy Prices in the Energy Community – State of Play and Recommendations for Reform; p. 3. Retrieved from https://www.energy-community.org/portal/page/portal/ENC_HOME/DOCS/1568177/0633975AB98C7B9CE053C92FA8C06338.PDF

Consequently, national markets are not equally competitive; neither is there much cross-border competition. Thus, we don't have an integrated and fully competitive electricity market that could offer new business opportunities, efficiency gains, price competition –as electricity is a homogeneous product–, higher standards of service and security of supply.

# III. The determinants of electricity price volatility.

Precisely, as aforementioned, the high volatility of electricity price constitutes the main concern of States when deciding whether to open or not the retail market to competition. They want to assure that all end-users will get access to electricity to ensure that they won't be victims of price fluctuation.

Thus, with price volatility being the main cause of retail market regulation, we ought to study the determinants of such fluctuation in prices. Importantly, we should add that even though price volatility is a common feature of energy markets (oil, gas and electricity markets), this is especially relevant for the latter[9].

For this purpose, we analyse the behaviour of demand and supply in the electricity market. On the one hand, electricity suppliers must ensure that end-users get electricity whenever they need it. To be clear, energy security is a State priority. However, we must take into account that electricity, due to its physical attributes, cannot be economically stored[10]. Therefore, electricity inventories do not exist. Indeed, this implies that there will be a real-time balancing between demand and supply, i.e., electricity is produced at the time it is consumed and so it will be instantaneously delivered (electricity flows at the speed of light). Consequently, electricity suppliers see demand variability as a challenge they have to deal with.

The only way electricity producers can adjust faster to changes in demand is by means of increasing the generation capacity or, at least, increasing the inventories of the fuel used to produce electricity. Nevertheless, some observations ought to be made regarding both strategies.

Firstly, investing in extra generation capacity that would rarely operate is extremely costly for firms. Thus, they generally prefer to increase the inventories of the fuel they use to generate electricity. However, storage costs shouldn't be overlooked. As a result, producers' generation capacity remains limited. Moreover, important to note is that the electricity supply cost curve is steep; in particular, it is "hockey stick" shaped[11]. That is, as the supply system gets closer to its capacity limit, the marginal generation costs escalate rapidly. Consequently, we expect periods of peak demand to be characterised by generation constraints and higher production costs.

---

[9] Energy Information Administration. US Department of Energy. (2002). Derivatives and Risk Management in the Petroleum, Natural Gas, and Electricity Industries; p. XIV. Retrieved from http://www.ksg.harvard.edu/hepg/Papers/DOE_Derivatives.risk.manage.electric_10-02.pdf

[10] Liu, M. and Wu, F.F. (2007). Risk management in a competitive electricity market. *Electrical Power & Energy Systems* 29; p. 690 and Homayoun Boroumand, R., Goutte, S., Porcher, S., and Porcher, T. (2015). Hedging strategies in energy markets: The case of electricity retailers. *Energy Economics* 51; p.503.

In addition, we should bear in mind other factors that can jeopardize an efficient electricity provision at reasonable prices to end-users. First, some plant outages can happen; hence reducing the supply. Second, in periods of peak demand, the reliability of the physical network can be undermined due to congestion of the network. Finally, it may be very difficult to ship supplies from areas where prices are low to areas where prices are high, due to the limited capability of the physical network connecting customers to suppliers[12]. On the other hand, some remarks ought to be made as regards with the resources companies use in order to generate electricity. In this sense, natural gas has been increasingly used[13]. Different reasons explain this: first, as electricity is produced when it's consumed (it cannot be cheaply stored in any meaningful quantity), the main cost of production becomes the cost of the fuel used in its generation. This includes both the market price of the resource and, for instance, the penalties that firms may have to pay when burning such fuel to generate electricity, due to the fact that it may cause a number of dioxide emissions that exceeds the cap set up by regulation[14]. In this respect, natural gas emits less carbon dioxide than coal or oil and it has become cheaper[15]. Second, unlike renewable resources such as solar and wind energy, natural gas allows constant supply and, with it, less price volatility. Moreover, power supply from natural gas can be adjusted much more quickly than power from other sources[16]. Notwithstanding the preceding advantages, it should be noted that natural gas is a resource whose cost could unexpectedly increase due to international political events (e.g. a political crisis with natural gas supply countries).

---

[11] Energy Information Administration. US Department of Energy. (2002). Derivatives and Risk Management in the Petroleum, Natural Gas, and Electricity Industries; p. 34. Retrieved from http://www.ksg.harvard.edu/hepg/Papers/DOE_Derivatives.risk.manage.electric_10-02.pdf

[12] Energy Information Administration. US Department of Energy. (2002). Derivatives and Risk Management in the Petroleum, Natural Gas, and Electricity Industries; p. 10. Retrieved from http://www.ksg.harvard.edu/hepg/Papers/DOE_Derivatives.risk.manage.electric_10-02.pdf

[13] Mack, I. M. (2014). *Energy trading and risk management: A practical approach to hedging, trading and portfolio diversification*. Hoboken: John Wiley & Sons; p. 14.

[14] Carmona, R., Fehr, M., and Hinz, J. (2009). Properly designed emissions trading schemes do work! p. 1. CE053C92FA8C06338.PDF

[15] Gas (n.d.) Greenpeace España. Retrieved from http://www.greenpeace.org/espana/es/Trabajamos-en/Frenar-el-cambio-climatico/Gas-natural/

[16] Mork, E. (2001). Emergence of financial markets for electricity: a European perspective. *Energy Policy*; p. 14.

Turning to the demand side, the first striking feature is that it's largely driven by weather conditions and especially temperature[17]. Furthermore, changes in demand use to happen at each hour within the same day[18]. This, combined with its inelasticity, draws up a stochastic pattern. To illustrate this, keep in mind that states use to fix retail prices, i.e., the retail price consumers see does not vary with the wholesale price of electricity. Therefore, if there is an increase in the demand (e.g. due to a change in weather conditions), this will be followed by an increase in wholesale prices (as suppliers face some constraints, according to what we stated before), but retail prices would not necessarily increase (as they are fixed). Thus, we do not expect consumers to respond to changes in wholesale prices, and so the demand will keep on increasing and so will the wholesale prices. On the other hand, when there is a decrease in the demand, the dynamic is the other way around, i.e., consumers will keep on decreasing the demand even if wholesale prices go down. The outcome is that consumers consume too much when supplies are stressed and too little when supplies are ample[19]. Furthermore, electricity cannot be substituted in the short term for other products.

Once we have pointed out the determinants of electricity price volatility, both from the supply and demand sides, we may consider to what extent derivatives and, in particular, options can be useful financial tools for managing price fluctuations.

---

[17] Benth, F. E. (2014). *Quantitative Energy Finance: Modeling, Pricing, and Hedging in Energy and Commodity Markets*. New York: Springer; p. 57.

[18] Homayoun Boroumand, R., Goutte, S., Porcher, S., and Porcher, T. (2015). Hedging strategies in energy markets: The case of electricity retailers. *Energy Economics* 51; p. 504.

[19] Energy Information Administration. US Department of Energy. (2002). Derivatives and Risk Management in the Petroleum, Natural Gas, and Electricity Industries; p. 10. Retrieved from http://www.ksg.harvard.edu/hepg/Papers/DOE_Derivatives.risk.manage.electric_10-02.pdf; p. XV.

**IV. Derivatives and options for managing electricity price volatility.**

With the deregulation of the electricity market, i.e., moving away from regulated slow-moving prices to deregulated fluctuating prices, a financial market emerged for this commodity, though it is not as significant as in the petroleum and natural gas industries[20].

In the financial market, as opposed to the physical market, no physical energy transactions are involved. This is a market where several financial instruments are provided (e.g. options) to offset ("hedge") the particular sources of risk in the electricity market (e.g. price volatility)[21].

To this end, as we anticipated, we are especially interested in ascertaining the usefulness of derivatives and, in particular, options as financial instruments with whom manage the high price volatility of electricity. In this sense, they may constitute an alternative to other risk management strategies which do not involve derivative contracts and that, sometimes, might not be adequate or affordable. For example, diversification – understood as investing in a variety of unrelated businesses – can be effective for reducing a firm's dependence on the performance of a particular industry (e.g. in our case, the electricity industry). However, diversification is expensive and therefore small and medium-sized firms cannot afford it. Moreover, it often fails because of the complexity of managing diverse businesses[22].

Turning to derivatives (and options), we provide a definition of them in order to find out why these financial instruments might be effective tools for mitigating exposure to electricity price risk.

In this sense, derivatives are financial instruments (contracts) that do not represent ownership rights in any asset but, rather, derive their value from the value of some underlying commodity -– in our case, electricity –. In addition, options are a specific kind of derivative contract that gives the buyer of the contract the right to buy (a call option) or sell (a put option) – in our case,

---

[20] Energy Information Administration. US Department of Energy. (2002). Derivatives and Risk Management in the Petroleum, Natural Gas, and Electricity Industries; p. 10. Retrieved from http://www.ksg.harvard.edu/hepg/Papers/DOE_Derivatives.risk.manage.electric_10-02.pdf; p. 39.

[21] Liu, M. and Wu, F.F. (2007). Risk management in a competitive electricity market. *Electrical Power & Energy Systems* 29; p. 691.

[22] Energy Information Administration. US Department of Energy. (2002). Derivatives and Risk Management in the Petroleum, Natural Gas, and Electricity Industries; p. 10. Retrieved from http://www.ksg.harvard.edu/hepg/Papers/DOE_Derivatives.risk.manage.electric_10-02.pdf; pp. 4 and 15.

electricity – at a specified price (the "strike price") over a specified period of time. Consequently, a call option sets a price ceiling, whereas a put option sets a floor price. Moreover, when producers fear that a current profitable margin (or spread between their input and output prices) will disappear, they can "lock" it by means of a spread option – in our case, a spark spread option[23] –.

At this point, the difference between American and European options is that the former allows the buyer to exercise his right at any time until the option expires, while European options can be exercised only at maturity. In any case, the buyer of the option pays for it in advance for the price protection they desire. Furthermore, we can expect the option's price to be more expensive when the terms are more favourable to the buyer and the longer its lifespan.

On the other hand, even though these options can be sold either on an exchange or on the over-the-counter (OTC) market, the majority of transactions occur in the OTC market[24]. In this sense, the OTC market offers the counterparties the possibility to negotiate option terms (e.g. price and maturity), whereas exchange-traded options are standard and inflexible, i.e., in our case, all options for electricity and a particular date are the same. Nevertheless, because OTC options are entered on a principal-to-principal basis (by means of direct contact between two parties or contact mediated by a broker), each counterparty is exposed to the credit risk of the opposite party. Overall, counterparties in the electricity financial market prioritize customization over liquidity and credit assurances[25].

Options are sold by speculators capable to absorb price fluctuations. They will do so as long as they believe that it's profitable for them. For this purpose, they exploit information inefficiencies, mispriced instruments and their higher risk capacity. Consequently, the presence of speculators, who try to benefit from the unwillingness of electricity suppliers to deal with price volatility, is crucial for the existence of financial electricity markets.

---

[23] Energy Information Administration. US Department of Energy. (2002). Derivatives and Risk Management in the Petroleum, Natural Gas, and Electricity Industries; p. 10. Retrieved from http://www.ksg.harvard.edu/hepg/Papers/DOE_Derivatives.risk.manage.electric_10-02.pdf; pp. 4 and 7.

[24] Deng, S.J. and Oren, S.S. (2006). Electricity derivatives and risk management. *Energy*; p.945.

[25] Energy Information Administration. US Department of Energy. (2002). Derivatives and Risk Management in the Petroleum, Natural Gas, and Electricity Industries; p. 10. Retrieved from http://www.ksg.harvard.edu/hepg/Papers/DOE_Derivatives.risk.manage.electric_10-02.pdf; pp. 45-48.

Furthermore, we should point out why electricity suppliers are so interested in covering against price fluctuations by means of options. First, the electricity industry is capital intensive, i.e., if firms want to enter into the market, they will have to face significant investments. Therefore, the will only do so as long as the expected generated cash overcomes the costs of the investment. In this sense, price volatility aggregates uncertainty to the cash-flows electric companies expect to receive. In turn, this implies that fewer firms will be willing to enter into the market, lessening competition and energy security. To illustrate this, keep in mind the negative shocks faced by retailers and generators of electricity[26].

On the one hand, retailers are in financial distress either when demand is high or when demand is low. In the first case, i.e., periods of peak demand, this is due to the fact that supply and demand's features entail an increase in wholesale prices, while retail prices are, in general, fixed. In the second case, i.e., periods of minimum demand, they face liquidity problems because the sales volume is too low to cover fixed costs.

On the other hand, generators are in financial distress when demand is low, since both the price and the volume are low (remember that consumers do not respond to changes in wholesale prices, due to the fact retail prices remain fixed).

Consequently, price volatility is associated with financial risks for electricity suppliers. Ultimately, this can end up with bankruptcies (and so lower energy supply). For this reason, we suggest that options can be welfare enhancing[27]. At least two reasons support it.

First, they allow perfect risk transfer from electricity suppliers to speculators. For instance, retailers usually have to deal with flexible consumer contracts, i.e., consumers can decide the volume of electricity they want to consume in each moment. Therefore, there can be a sudden increase in the demand. In this case, retailers who purchase call options will know which is the maximum price (ceiling price) they will have to pay for additional electricity; the speculator will pay the difference between the market price and the "strike price".

Another example is the case electricity producers, who want to assure profitable margins by means of spark spread options. These cross-commodity options pay out the difference between the price of electricity sold by the generator and the price of the fuels used to generate it[28]. Thus, they allow electricity producers to "lock" a specific margin in the event of lower demand.

---

[26] Willems, B. and Morbee, J. (2010). Market completeness: How options affect hedging and investments in the electricity sector. *Energy Economics*; p.792.

[27] Willems, B. and Morbee, J. (2010). Market completeness: How options affect hedging and investments in the electricity sector. *Energy Economics*; p.786.

Second, if electricity suppliers can transfer the financial risks associated with price volatility, they will face less financial distress and they will enjoy more cash-flow stability. Therefore, electricity suppliers will have more incentives to invest in this industry, fostering competition in the market. Moreover, this might bring about more efficient activities at lower prices.

---

[28] Deng, S.J. and Oren, S.S. (2006). Electricity derivatives and risk management. *Energy*; p. 945.

## V. . The European Commission's further deregulation of the energy market (focus on the electricity market).

We have already mentioned that the goal of the European directives in the electricity sector was to achieve a single and competitive energy market. To this end, they encouraged Member States to fully deregulate their domestic electricity markets. Nonetheless, and even though there has been a trend towards less public ownership of electricity assets and less price regulation, the reality is that in 2013 around 15 EU Member States still regulated end-user prices for electricity[29]; the liberalisation and integration of the retail energy market are still not complete[30].

For many years, these States have based their regulation of retail prices on the prerogative attributed by the European directives. According to this right, they can impose on undertakings operating in the electricity sector the price of supplies if consumers' access to electricity is in danger. Precisely, fixing retail prices is a method to prevent vulnerable consumers from being deprived of electricity consumption in periods of peak demand (when the price rises). Moreover, it should be added that consumers prefer to stabilize or smooth their budgets. In this sense, avoiding price fluctuation is a way to do so.

Nevertheless, the European Commission has recently restated the advantages of deregulating the retail market in his Communication of 15th July 2015[31]. On the one hand, it will foster investments in low-cost supply capacities. This is due to the fact that electricity is a homogenous product, and so it's highly subject to price competition. Therefore, consumers will benefit from lower prices; they will have the right to shop around for the best energy deal.

On the other hand, it will improve the real-time balancing between demand and supply, since there will be less demand peaks. This is because deregulation of prices increases demand's response or elasticity. That is, in periods of peak demand, consumers will experience increases in retail prices and so they will adjust their consumption as prices evolve. In turn, this implies that there will be less price volatility in the wholesale market. Furthermore, the European Commission has proposed to further improve demand's response by means of, first, lower network charges for consumers who reduce their consumption when the network is congested (though consumers unable to shift their demand must not be penalized) and, second, via provision of information regarding the costs and levels of energy consumption to end-users (so as they can change their habits).

---

[29]European Parliament (2016). A New Deal for energy consumers; p. 4. Retrieved from http://www.europarl.europa.eu/RegData/etudes/BRIE/2016/573896/EPRS_BRI(2016)573896_EN.pdf

[30] Committee on Industry, Research and Energy. European Parliament. Delivering a new deal for energy consumers (2016); p. 13. Retrieved from http://www.europarl.europa.eu/sides/getDoc.do?pubRef=-//EP//TEXT+REPORT+A8-2016-0161+0+DOC+XML+V0//EN

[31] Communication from the Commission to the European Parliament, the Council, the European Economic and Social Committee and the Committee of the Regions (2015). Delivering a New Deal for Energy Consumers.

However, two conditions must hold in order to ensure competition in the retail market and, at the same time, that vulnerable consumers will not be deprived from electricity. As regards with truly competition among suppliers, we need both many suppliers and the possibility to switch between them easily and quickly (e.g. being provided with transparent and directly comparable offers and not being hampered with switching fees).

Finally, and turning to vulnerable consumers, who may be at the wrong end of price fluctuation, the European Commission has acknowledged the need to protect them. Nevertheless, it doesn't believe that this has to be done by means of fixing retail prices permanently. Accordingly, Member States should take care of vulnerable consumers through national social welfare systems, since vulnerability of consumers typically does not refer to energy consumption only.

Moreover, if Member States want to use their prerogative to impose the price of supplies on undertakings operating in the electricity sector (so as to guarantee electricity access to vulnerable consumers), they ought to do so according to the ruling of the Court of Justice. That is, imposition of supply prices needs to be limited in duration and should not go beyond what is necessary to achieve the objective pursued. For instance, prices could be regulated in order to limit the impact of the increase in the price of fuel that would in the absence of intervention have had a major impact on the sale price offered to final customers. The regulation, however, should be limited only to the price component influenced by the specific circumstances, but not the final end-user's price[32].

---

[32] Energy Community. (2012). Regulated Energy Prices in the Energy Community – State of Play and Recommendations for Reform; pp. 5-6. Retrieved from https://www.energy-community.org/portal/page/portal/ENC_HOME/DOCS/1568177/0633975AB98C7B9CE053C92FA8C06338.PDF

# VI. Introduction to options with random strikes and spread options.

In this paper we study options with random strikes, such as two-asset spread options. So let us consider a spread option with a pay-off given by:

$$S_3 = (S_1 - S_2 - K)_+, (1.1)$$

where $S_1$ and $S_2$ denote the prices of two underlying assets and have the following form:

$$S_1 = S_t * e^{r - \frac{\sigma_1^2}{2}T + \sigma_1 * \sqrt{T} * W} \quad (1.2.1)$$

$$S_2 = S_t' * e^{r - \frac{\sigma_2^2}{2}T + \sigma_2 * \sqrt{T} * B}, \quad (1.2.2)$$

where $S_t$ and $S_t'$ are the initial prices for the underlying assets, $\sigma_1$ and $\sigma_2$ denote the volatility parameters of $S_t$ and $S_t'$ respectively and $\rho$ is the correlation. $K$ is a constant called the strike price, $T$ is the time parameter and $r$ is our risk-free interest rate, which is $r = 0$.

In this case, after describing the energy markets and the role played by spread options in these, we will introduce different option pricing solutions.

This introductory section is followed by three sections that delve into the different option pricing techniques. In the first section we will introduce the Margrabe's formula used to price the spread options in the case $K = 0$. The second section will address the issue of lack of an analytical formula, when $K$ is not 0. In this case an approximation exists, called Kirk's Approximation.

In the third section, we show the recent developments of the option pricing approximation and most importantly the Modified Kirk's Approximation, which gives better results the Kirk's formula, for correlations that are close to 1. Finally in the fourth section we will develop a program that will help study numerically the Modified Kirk's Approximation.

## VII. Margrabe's formulation and Kirk's Approximation

William Margrabe, in his seminal work[33][34], published in 1978, exposed a formula that gave an explicit solution for calculating the option price when $K$ is 0, which is a constant and can be called as the strike price. This way, Margrabe developed an equation to price an option to exchange one risky asset for another. The general Magrabe's formula follows as this:

$$C\left(S_t, S_t^{'}, t\right) = S_t N(d_1) - S_t^{'} N(d_2) \quad (2.1)$$

$$d_1 = \frac{\ln(S_t/S_t^{'})}{\sigma} + \frac{\sigma}{2} \quad (2.2.1)$$

$$d_2 = \frac{\ln(S_t/S_t^{'})}{\sigma} - \frac{\sigma}{2} \quad (2.2.2)$$

$$\sigma = \sqrt{\sigma_1^2 - 2\rho\sigma_1\sigma_2 + \sigma_2^2}, \quad (2.3)$$

where $N(\cdot)$ denotes the cumulative standard normal density function, $S_t$ and $S_t^{'}$ are the initial prices for the underlying assets, $\sigma$ is the volatility parameter of Magrabe's Formula, $\sigma_1$ and $\sigma_2$ denote the volatility parameters of $S_t$ and $S_t^{'}$ respectively and $\rho$ is the correlation.

As we have said earlier, Margrabe's formula is an analytical solution for the case when our $K$ is 0, but in the case when $K$ is different from 0 no explicit solution exists. In the latter case, when our $K$ is different from 0, there exist some approximations, being Kirk's formula the most popular one.

Ewan Kirk[35], in 1995, introduced a simple but powerful approximation technique for spread option valuation that we know today as the Kirk's formula. Its main advantage is that it can be applied to spread option pricing when our $K$ is small.

Also, another important advantage is that it doesn't bear the computational costs that bear the Monte Carlo simulations. As we will see it is a modification of the original Black-Scholes formula, with an introduction of a different expression of the parameter $\sigma$, which we will refer to as $a_t$. The Kirk's Approximation can be defined as:

$$a_t = \sqrt{\sigma_1^2 - 2\rho\sigma_1\sigma_2 \frac{S_t'}{S_t'+K} + \sigma_2^2 \frac{(S_t')^2}{(S_t'+K)^2}} \quad (3.1)$$

$$S = \frac{S_t}{(S_t'+K)} \quad (3.2)$$

$$d_1 = \frac{\ln(S) + \left(\frac{a_t}{2}\right)T}{a_t\sqrt{T}} \quad (3.3.1)$$

$$d_2 = d_1 - a_t\sqrt{T} \quad (3.3.2)$$

$$C_{BS} = e^{-rT} S_t N(d_1) - S_t' N(d_2), \quad (3.4)$$

where $S_t$ and $S_t'$ are the initial prices for the underlying assets, $a_t$ denotes the volatility parameter of Kirk's Approximation, $\sigma_1$ and $\sigma_2$ denote the volatility parameters of $S_t$ and $S_t'$ respectively, $\rho$ is the correlation, $K$ is a constant called the strike price, $T$ is the time parameter, which is $T = 1$, $N(\cdot)$ denotes the cumulative standard normal density function and $r$ is our risk-free interest rate, which is $r = 0$.

---

[33] Margrabe, W. (1978). The Value Of An Option To Exchange One Asset For Another. *The Journal of Finance, 33*(1), 177-186. doi:10.1111/j.1540-6261.1978.tb03397.x. Retrieved from http://www.stat.nus.edu.sg/~stalimtw/MFE5010/PDF/margrabe1978.pdf.

[34] Hull, J. (2014). *Options, futures, and other derivatives* (9th ed.). Upper Saddle River, NJ: Pearson/Prentice Hall; pp. 611-612

[35] Kirk, E. (1995). Correlation in the Energy Markets. In, Robert Jameson (Ed.), *Managing Energy Price Risk* (pp. 71-78). London: Risk Publications and Enron Capital & Trade Resources.

. As we commented previously, the Kirk's Approximation is a useful tool and can be applied to spread option pricing when our $K$ is small or takes a 0 value (because in a sense, Margrabe's formulation is special case of Kirk's Approximation, matching its results with those of Margrabe's). The approximation starts to fail and problems arise when $\rho$ is close to 1.

In the energy markets, it is very usual to encounter situations where the correlations are very high, being typical examples the price of crude oil and the price of refined oil, the price of natural gas and the price of electric power. Hence it is imperious to find an improvement of the approximation in the cases when $\rho$ is close to 1, which we will discuss in the next section.

## VIII. Modified Kirk's Approximation.

Elisa Alòs Alcalde and Jorge Alberto León Vázquez, in 2016, introduced a modification[36] for the Kirk's Approximation formula. The authors, based on Malliavin calculus, arrive to a result that can be applied to spread option pricing when $K$ is different from 0 and more importantly when $\rho$ is close to 1, improving this way Kirk's Approximation formula. The Modified Kirk's Approximation is defined this way:

$$a_t = \sqrt{\sigma_1^2 - 2\rho\sigma_1\sigma_2 \frac{S_t'}{S_t'+K} + \sigma_2^2 \frac{(S_t')^2}{(S_t'+K)^2}} \quad (4.1)$$

$$X_t = \ln(S_t) \quad (4.2.1)$$

$$x_t^* = \ln(S_t' - K) \quad (4.2.2)$$

$$\hat{I}_t(X_t) = \sqrt{a_t^2} + \frac{1}{2}\left(\sigma_2\left(\frac{S_t'}{S_t'+K}\right) - \rho\sigma_1\right)^2 \frac{1}{(\sqrt{a_t^2})^3} \sigma_2^2 \frac{S_t'K}{(S_t'+K)^2}(X_t - x_t^*) \quad (4.3)$$

$$S = \frac{S_t}{(S_t'+K)} \quad (4.4)$$

$$d_1 = \frac{\ln(S) + \left(\frac{\hat{I}_t}{2}\right)T}{\hat{I}_t\sqrt{T}} \quad (4.5.1)$$

$$d_2 = d_1 - \hat{I}_t\sqrt{T} \quad (4.5.2)$$

$$C_{BS} = e^{-rT} S_t N(d_1) - S_t' N(d_2), \quad (4.6)$$

where $S_t$ and $S_t'$ are the initial prices for the underlying assets, $a_t$ denotes the volatility parameter for the Modified Kirk's Approximation, $\sigma_1$ and $\sigma_2$ denote the volatility parameters of $S_t$ and $S_t'$ respectively, $\rho$ is the correlation, and $K$ is a constant called the strike price.

Also, the $\hat{I}_t$ is an approximation of the implied volatility that takes into account its skew, which in turn depends on the $X_t$, the $T$ is the time parameter, which is $T = 1$, $N(\cdot)$ denotes the cumulative standard normal density function and $r$ is our risk-free interest rate, which is $r = 0$. By means of modifying the volatility parameter in the Kirk's Approximation, the authors have achieved an approximation that is still easy and simple to apply, but is more accurate. This modification of the Kirk's Approximation formula is a highly improved technique that can be applied to spread option pricing when our when $K$ is different from 0 and more importantly when $\rho$ is close to 1. The improvement yields significantly lower errors than the original Kirk's formula, but in the paper there are few numerical examples that study the goodness of fit of the new formula, for that reason we will address this issue in the next section.

---

[36] Alòs, E., & León, J. A. (2015). On the short-maturity behaviour of the implied volatility skew for random strike options and applications to option pricing approximation. *Quantitative Finance, 16*(1), 31-42. doi:10.1080/14697688.2015.1013499. Retrieved from http://dx.doi.org/10.1080/14697688.2015.1013499.

# IX. Modified Kirk's Approximation's numerical simulations.

In this section we will widen the scope of the paper discussed in the last section, for that reason we have developed a program, using the mathematical and statistical language R, which will allow us to study the goodness of fit of the Modified Kirk's Approximation in different scenarios.

With this objective in mind, first of all, because the paper published by Alòs and León had relatively few trials (namely, $n = 1,000,000$ trials) we have performed a Monte Carlo simulation to simulate the spreads, with $n = 5,000,000$ trials, in order to obtain much more reliable and robust results. We have to explain that the Monte Carlo simulations[37] are generally numerically more efficient than the other procedures when there are three or more stochastic variables; because the computational time to perform a Monte Carlo simulation increases approximately linearly with the number of variables that we add, while for other procedures it increases exponentially.

Besides this, there are three more advantages to a Monte Carlo simulation, namely, it can calculate the standard error for the estimates, it can accommodate pay-offs and stochastic processes of varying complexity, and it can also be used when the pay-off depends on a function of the whole path followed by a variable and not just its terminal value.

In order to perform this, first we simulated two Brownian motions, with $n = 5,000,000$ trials for each value, with mean 0 and a standard deviation of 1.

```
n = (5000000)
W = rnorm(n,mean=0,sd=1)
Z = rnorm(n,mean=0,sd=1)
```

We also chose $\rho = 0.999$, and after obtaining two different simulations our Brownians, namely $X$ and $Z$, we were able to obtain a third value, $B$, in such way that $X$ and $B$, would be correlated, with the following formula:

$$B = \rho W + \left(\sqrt{1 - (\rho^2)}\right)Z \qquad (5.1)$$

---

[37] Hull, J. (2014). *Options, futures, and other derivatives* (9th ed.). Upper Saddle River, NJ: Pearson/Prentice Hall; pp. 475-476.

In the next step we simulated our spreads with each value, wrote the pay-off form and found the mean of the pay-off for a specific scenario, which is $K = 5$, time is $T = 0.5$, interest rate $r = 0$, $\sigma_1 = 0.3$, $\sigma_2 = 0.2$ and the initial value of the asset $S_0 = 100$. The following piece of code achieves this purpose.

```
##Here we introduce the inputs for S1
##inputs:
r=0; sigma1 =0.3; T=0.5; n=5000000; S0_1=100;
S1 = S0_1 * exp((r -(sigma1^2)/2)  * T + sigma1 * sqrt(T) * W)
##Here we introduce the inputs for S2
##inputs:
r=0; sigma2 =0.2; T=0.5; n=5000000; S0_2=100;
S2 = S0_2 * exp((r -(sigma2^2)/2) * T + sigma2 * sqrt(T) * B)
##Here we calculate max(S^1(T) - S^2(T) - 0, 0)
S3 = pmax(S1 - S2 - K, 0)
S3_K0mean = mean(S3)
S3_K0mean
```

After, we performed a method of the reduction of the variance. These procedures of reduction of the variance are very important in this case, in order to obtain much more reliable and robust results, than those presented in the paper by Alòs and León. We have to remember, that if we simulate the stochastic processes for the variables underlying a derivative, as indicated in the equations (1.2.1) and (1.2.2) we will need a very large number of trials to estimate the value of the spread options with accuracy. A problem arises, which is the very high computational cost. To decrease and save computational time, we have performed a method of reduction of the variance.

This was achieved by the antithetic variables technique[38]. This involves obtaining two values of the derivative. The first value $S_3$ is calculated as described above; the second value $S_3'$ is calculated by changing the sign of all the random samples from the standard normal distributions. So if $X$ and $B$ are the samples to calculate values $S_1$ and $S_2$, then this same sample is used to calculate $S_1^{'}$ and $S_2^{'}$. In the way the sample value of the derivative calculated from a simulation trial is the average of $S_3$ and $S_3^{'}$.

---

[38] Hull, J. (2014). *Options, futures, and other derivatives* (9th ed.). Upper Saddle River, NJ: Pearson/Prentice Hall; pp. 475-476.

Theoretically, this performs exceptionally well, because we know when one value is above the true value, the other is below, and this works the other way too. We can denote $\bar{S}$ as the average of $S_3$ and $S_3^{'}$:

$$\bar{S} = \frac{S_3 + S_3^{'}}{2} \qquad (5.2)$$

After we can obtain the final estimate of the value of the derivative with the average of the $\bar{S}$. Finally we can obtain the standard error of the estimate, which has the following form:

$$\frac{\bar{\omega}}{\sqrt{M}}, \qquad (5.3)$$

where $\bar{\omega}$ is the standard deviation of the $\bar{S}$ and $M$ is the number of simulation trials (the numerb of pairs of values calculated). This techniques allows us to obtain a standard error calculated using $2M$ random trials. So after obtaining the antithetic variables, we repeated the above steps, which included the simulation of our spreads with each value, writing the pay-off form and finding the mean of the pay-off in the case when the $K = 0$.

```
##Here we calculate max(S^1(T) - S^2(T) - 0, 0)
invS3 = pmax(invS1 - invS2 - K, 0)
invS3_K0mean = mean(invS3)
invS3_K0mean
```

The final step, involved the calculation the mean of our obtained spreads, observing effectively a drop in our standard deviation. The following R code achieves this result.

```
S3_total_mean = (S3_K0mean + invS3_K0mean) / 2
S3_total_mean
```

Once obtained the results we continue to the next step, which is performing the Kirk's Approximation commented above, in the following piece of R code.

```
##Here we add Kirk's approximation for spread option prices

##Here we calculate our volatility

a_kirk = sqrt( (sigma1)^2 - 2*rho*sigma1*sigma2 * (S0_2/(S0_2 + K)) + (sigma2)^2 *
((S0_2/(S0_2 + K))^2) )

##Here we calculate our d1 and d2

S = (S0_1 / (S0_2 + K))

d1 = (log(S, base = exp(1)) + 1/2 * (a_kirk^2) * T) / ( a_kirk * (sqrt(T)) )

d2 = d1 - a_kirk * sqrt(T)

##Here we use the above calculations to approximate the call spread using Kirk's formula

N_d1 = pnorm(d1,mean=0,sd=1)

N_d2 = pnorm(d2,mean=0,sd=1)

C_kirk = (exp(-r*T)) * ( (S0_1*N_d1) - ((S0_2 + K) * N_d2) )

C_kirk
```

After obtaining our results of the approximation, now we can finally do the Modified Kirk's Approximation that we discussed in the third section, with the following code in R.

```
##Here we add Modified Kirk's approximation for spread option prices

##Here we calculate our volatility

a_kirk = sqrt( (sigma1)^2 - 2*rho*sigma1*sigma2 * (S0_2/(S0_2 + K)) + (sigma2)^2 *
((S0_2/(S0_2 + K))^2) )

X_t = log(S0_1, base = exp(1))

x_aster = log((S0_2 + K), base = exp(1))

I_t = sqrt(a_kirk^2) + 1/2 * (( (sigma2 * S0_2/(S0_2 + K)) - rho* sigma1)^2 ) *

( 1 / ((sqrt(a_kirk^2))^3) ) *(sigma2^2) * ( (S0_2 * K) / ((S0_2 + K)^2) ) * (X_t - x_aster)

##Here we calculate our d1 and d2

S = (S0_1 / (S0_2 + K))

d1_modif = (log(S, base = exp(1)) + 1/2 * (I_t^2) * T) / (I_t * (sqrt(T)) )

d2_modif = d1_modif - I_t * sqrt(T)
```

```
##Here we use the above calculations to approximate the call spread using Modified Kirk's formula

N_d1_modif = pnorm(d1_modif,mean=0,sd=1)

N_d2_modif = pnorm(d2_modif,mean=0,sd=1)

C_kirk_modif = (exp(-r*T)) * ( (S0_1*N_d1_modif) - ((S0_2 + K) * N_d2_modif) )

C_kirk_modif
```

Here we include the confidence interval tables with some examples to illustrate that our numerical method has worked perfectly. The confidence intervals were obtained as usually and the following are the results:

| Confidence Intervals | $\rho = 0.9$ | $\rho = 0.999$ |
|---|---|---|
| $K = 5$ | (2.357551, 2.363762) | (1.273913, 1.278092) |
| $K = 10$ | (1.26478, 1.269644) | (0.5398617, 0.5427516) |

Here we clearly can observe that the intervals are too narrow, giving us a clear indication that our numerical method performed effectively. If we compare these results with the results we obtained for Kirk's Approximation and Modified Kirk's Approximation we can confirm clearly that our numerical method performed positively. Here we present some examples of the Kirk's Approximation results:

| Kirk's Approximation | $\rho = 0.9$ | $\rho = 0.999$ |
|---|---|---|
| $K = 5$ | 2.3647228 | 1.2862590 |
| $K = 10$ | 1.2745318 | 0.5615868 |

And also some examples of the Modified Kirk's Approximation results:

| Modified Kirk's Approximation | $\rho = 0.9$ | $\rho = 0.999$ |
|---|---|---|
| $K = 5$ | 2.3626873 | 1.27686463 |
| $K = 10$ | 1.2681347 | 0.54140923 |

The results presented here indicate us that in the cases when correlations are small, the Kirk's Approximation and the Modified Kirk's Approximation fall into the confidence intervals presented above, because at this level the Kirk's results are consistent. But as our correlations grow we can observe that Kirk's Approximation fails and falls out of the confidence intervals and only the Modified Kirk's Approximation stay within the confidence intervals, indicating us the reliability of this new method.

The following step involved obtaining the errors of the two approximations in relation to our original Monte Carlo simulation, with the following code.

```
##Here we calculate the errors of the Kirk and Modified Kirk's formulas
error_Kirk = ((C_kirk * 100) / S3_total_mean) - 100
error_Kirk
error_Kirk_modif = ((C_kirk_modif * 100) / S3_total_mean) - 100
error_Kirk_modif
```

As we stated previously, the above simulations were performed with a $K = 5$ and with a $\rho = 0.999$.

Being our main objective to widen the numerical examples to be able to study more thoroughly the results of the Modified Kirk's Approximation, we replicated the process described above with for a wide range of values of $K$ and $\rho$. We chose a range of $K$ from 0 to 20 and for $\rho$ we chose 5 values that are 0.80, 0.85, 0.90, 0.95 and 0.999.

Once obtained the results of the simulations, first we calculated the errors produced by the Kirk's Approximation in relation to the Monte Carlo simulation and secondly we computed the errors produced by the Modified Kirk's Approximation in relation to the Monte Carlo simulation. The following figures of the plots describe our results.

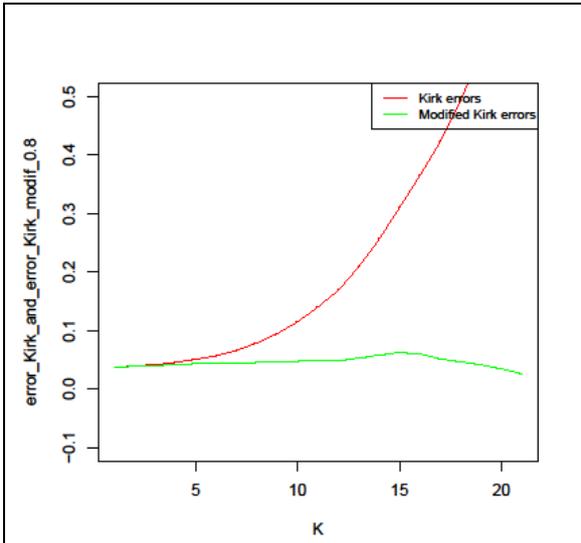

Figure 1. Here $\rho = 0.80$. The errors are in % terms of the benchmark (Monte-Carlo simulations).

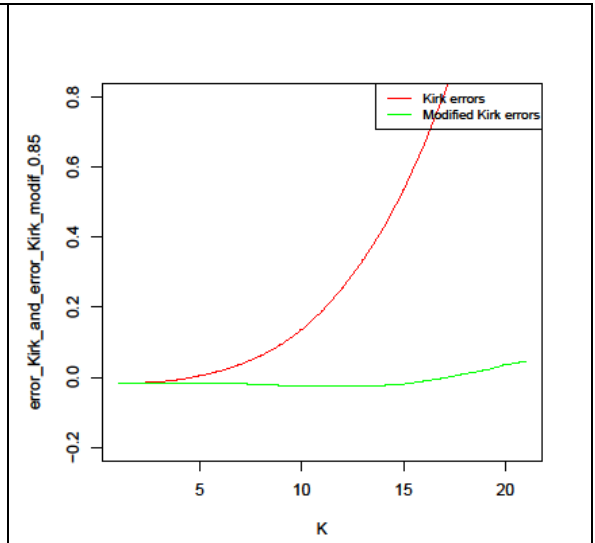

Figure 2. Here $\rho = 0.85$. The errors are in % terms of the benchmark (Monte-Carlo simulations).

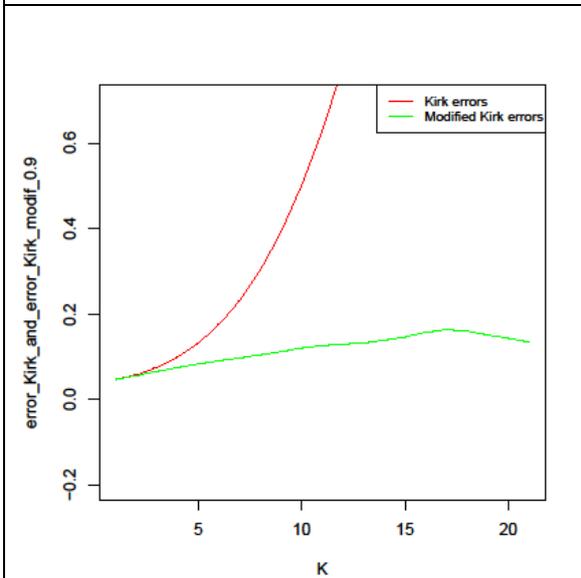

Figure 3. Here $\rho = 0.90$. The errors are in % terms of the benchmark (Monte-Carlo simulations).

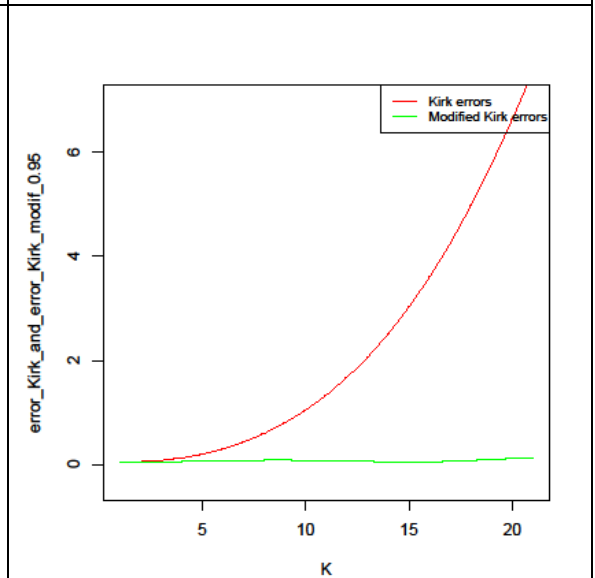

Figure 4. Here $\rho = 0.95$. The errors are in % terms of the benchmark (Monte-Carlo simulations).

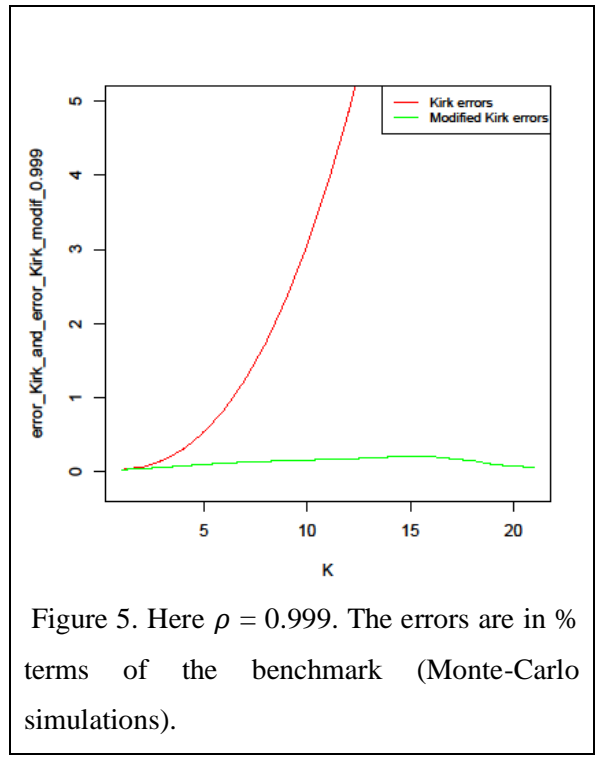

Figure 5. Here $\rho = 0.999$. The errors are in % terms of the benchmark (Monte-Carlo simulations).

As we can observe in the above figures, for example in Figure 5, with each increment in the value of $\rho$ and in the value of $K$ the errors of the Kirk's Approximation start taking an exponential shape. The contrary happened to the Modified Kirk's Approximation, because we can clearly devise that even if the values of $\rho$ and $K$ surge, the errors of the approximation in relation to the Monte Carlo simulations remain constant, except in the case of Figure 1.

This is one of the main advantages of the Modified Kirk's Approximation, because it allows, like the Kirk's Approximation, a fast computation of the spreads, without the great computational cost that can impose a Monte Carlo simulation with an important number of trials.

As we noted earlier, the previous computations were made, using a value of $T = 0.5$. To achieve a more complete image of our performance we have also performed the same set of calculations, but with different values of $T$. We chose 5 values that are 0.1, 0.2, 0.3, 0.4 and 0.5. The following figures of the surface plots would describe our results.

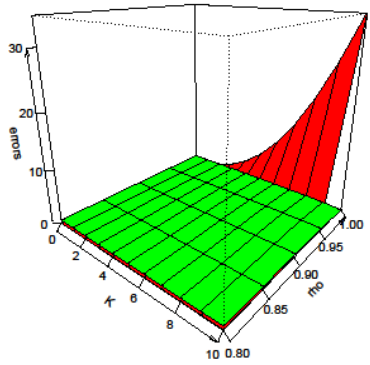

Figure 6. Here $T = 0.1$. The errors are in % terms of the benchmark (Monte-Carlo simulations).

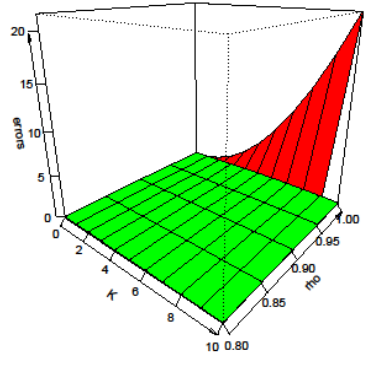

Figure 7. Here $T = 0.2$. The errors are in % terms of the benchmark (Monte-Carlo simulations).

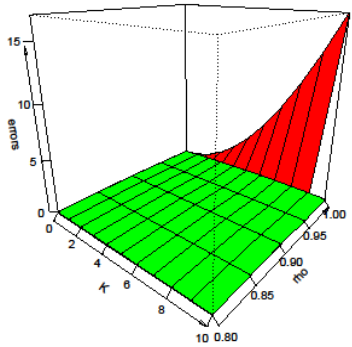

Figure 8. Here $T = 0.3$. The errors are in % terms of the benchmark (Monte-Carlo simulations).

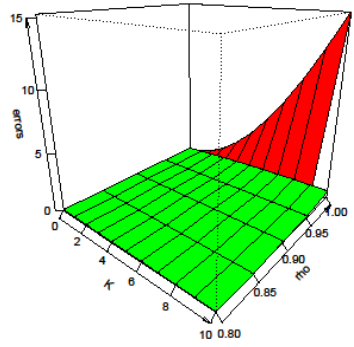

Figure 9. Here $T = 0.4$. The errors are in % terms of the benchmark (Monte-Carlo simulations).

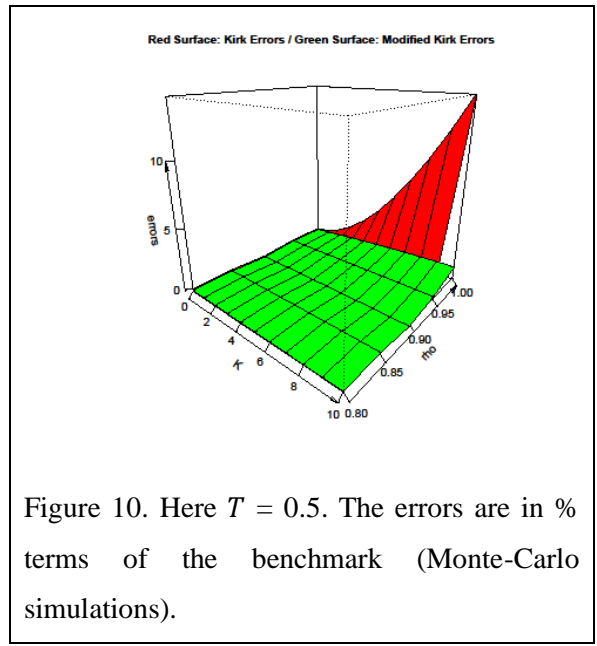

Figure 10. Here $T = 0.5$. The errors are in % terms of the benchmark (Monte-Carlo simulations).

As in our previous cases, we can clearly observe in the figures of the surface plots, with each increment in the value of $T$, the values of $\rho$, and in the values of $K$ the errors of the Kirk's Approximation start spiking and ultimately take an exponential shape. As before happened to the Modified Kirk's Approximation, because we can clearly observe that even if the values of $T$, $\rho$, and $K$ surge, the errors of the approximation in relation to the Monte Carlo simulations remain constant.

# X. Conclusions.

After reviewing the relatively young energy markets (and especially the electricity markets) and its regulatory framework we can ascertain that these new frontiers need innovative methods to manage the energy price risk. The two-asset and three-asset spread options are a natural choice, given that they enable the possibility to hedge the risk with a relatively low cost. We have seen that the Kirk's Approximation of the pricing of these options was a great breakthrough for financial mathematics field, because of its simplicity and accuracy. Alas, it was flawed in the cases of very high correlations, which is a very natural assumption in the energy markets. Hence, more advanced and innovative methods are needed to manage the new uncertainties posed by these new products. For that reason, the Modification of the Kirk's Formula is a great answer to that need, because by modifying the volatility parameters of the original formula posed by Kirk, the authors have achieved a simple, easy to apply and accurate solution to approximate spread option prices with a high correlation parameter.

This paper's particular task was reviewing and assessing the goodness of fit of this new method by performing more simulations than those presented in the original paper by Alòs and León[39], and by studying more scenarios. The simulations performed exceptionally well in the cases with a high correlation parameter and the results presented here ascertain that. We can clearly see that the Modified Kirk's Approximation method greatly reduces the errors in percentage terms of the Monte-Carlo simulation, unlike the Kirk's original method. Moreover the confidence intervals also show the goodness of fit of this new method, giving us a second confirmation, besides the figures presented in the paper, that the method is extremely accurate.

To say that this is the end of the road would be simple, simplistic and, even worse anachronistic. As the financial markets evolve and become even more sophisticated, new markets emerge and innovative products are created, more than ever before, advanced methods are needed to progress the field of quantitative finance and to push the boundaries even more.

---

[39] Alòs, E., & León, J. A. (2015). On the short-maturity behaviour of the implied volatility skew for random strike options and applications to option pricing approximation. *Quantitative Finance, 16*(1), 31-42. doi:10.1080/14697688.2015.1013499. Retrieved from http://dx.doi.org/10.1080/14697688.2015.1013499.

# XI. References.

Alòs, E., & León, J. A. (2015). On the short-maturity behaviour of the implied volatility skew for random strike options and applications to option pricing approximation. *Quantitative Finance, 16*(1), 31-42. doi:10.1080/14697688.2015.1013499. Retrieved from http://dx.doi.org/10.1080/14697688.2015.1013499.

Benth, F. E. (2014). *Quantitative Energy Finance: Modeling, Pricing, and Hedging in Energy and Commodity Markets*. New York: Springer; pp. 41-83.

Carmona, R. and Durrleman, V. (2003). Pricing and Hedging. Spread Options. *Society for Industrial and Applied Mathematics (SIAM) Review*; pp. 627-685.

Carmona, R., Fehr, M., and Hinz, J. (2009). Properly designed emissions trading schemes do work! pp. 1-7. CE053C92FA8C06338.PDF

Committee on Industry, Research and Energy. European Parliament. Delivering a new deal for energy consumers (2016); pp. 1-17. Retrieved from http://www.europarl.europa.eu/sides/getDoc.do?pubRef=-//EP//TEXT+REPORT+A8-2016-0161+0+DOC+XML+V0//EN

Communication from the Commission to the European Parliament, the Council, the European Economic and Social Committee and the Committee of the Regions (2015). Delivering a New Deal for Energy Consumers; pp. 1-10. Retrieved from https://ec.europa.eu/energy/sites/ener/files/documents/1_EN_ACT_part1_v8.pdf

Deng, S.J. and Oren, S.S. (2006). Electricity derivatives and risk management. *Energy*; pp. 940-953.

Directive 2003/54/EC of the European Parliament and of the Council of 26 June 2003 concerning common rules for the internal market in electricity and repealing Directive 96/92/EC.

Directive 2009/72/EC of the European Parliament and of the Council of 13 July 2009 concerning common rules for the internal market in electricity and repealing Directive 2003/54/EC.

Directive 96/92/EC of the European Parliament and of the Council of 19 December 1996 concerning common rules for the internal market in electricity.

Dorsman, A. B. (2011). *Financial aspects in energy: A European perspective*. Heidelberg: Springer; pp. 11-32.

Energy Community. (2012). Regulated Energy Prices in the Energy Community– State of Play and Recommendations for Reform; pp. 1-36. Retrieved from https://www.energy-community.org/portal/page/portal/ENC_HOME/DOCS/1568177/0633975AB98C7B9

Energy Information Administration. US Department of Energy. (2002). Derivatives and Risk Management in the Petroleum, Natural Gas, and Electricity Industries; pp. IX-69. Retrieved from http://www.ksg.harvard.edu/hepg/Papers/DOE_Derivatives.risk.manage.electric_10-02.pdf

European Parliament (2016). A New Deal for energy consumers; pp. 1-8. Retrieved from http://www.europarl.europa.eu/RegData/etudes/BRIE/2016/573896/EPRS_BRI(2016)573896_EN.pdf

Gas (n.d.) Greenpeace España. Retrieved from http://www.greenpeace.org/espana/es/Trabajamos-en/Frenar-el-cambio-climatico/Gas-natural/

Homayoun Boroumand, R., Goutte, S., Porcher, S., and Porcher, T. (2015). Hedging strategies in energy markets: The case of electricity retailers. *Energy Economics* 51; pp. 503-509.

Hull, J. (2014). *Options, futures, and other derivatives* (9$^{th}$ ed.). Upper Saddle River, NJ: Pearson/Prentice Hall; pp. 611-612 and 475-476.

Kirk, E. (1995). Correlation in the Energy Markets. In, Robert Jameson (Ed.), *Managing Energy Price Risk* (pp. 71-78). London: Risk Publications and Enron Capital & Trade Resources.

Liu, M. and Wu, F.F. (2007). Risk management in a competitive electricity market. *Electrical Power & Energy Systems* 29; pp. 690-697.

Mack, I. M. (2014). *Energy trading and risk management: A practical approach to hedging, trading and portfolio diversification*. Hoboken: John Wiley & Sons; pp. 1-17.

Margrabe, W. (1978). The Value Of An Option To Exchange One Asset For Another. *The Journal of Finance, 33*(1), 177-186. doi:10.1111/j.1540-6261.1978.tb03397.x. Retrieved from http://www.stat.nus.edu.sg/~stalimtw/MFE5010/PDF/margrabe1978.pdf.

Mork, E. (2001). Emergence of financial markets for electricity: a European perspective. *Energy Policy*; pp. 7-15.

Pillsbury Winthrop Shaw Pittman LLP. (2010). Electricity, Oil and Gas Regulation in the United States; pp. 188-198. Retrieved from http://www.ourenergypolicy.org/wp-content/uploads/2013/08/ElectricityOilandGasRegulationintheUnitedStates.pdf

Vehviläinen, I. and Keppo, J. (2003). Managing electricity market price risk. *European Journal of Operational Research*; pp.136-147.

Weron, R. (2000). Energy price risk management. *Physica A*; pp. 127-134.

Willems, B. and Morbee, J. (2010). Market completeness: How options affect hedging and investments in the electricity sector. *Energy Economics*; pp. 786-795.

Wilmott, P. (2001). *Paul Wilmott introduces quantitative finance*. Chichester: John Wiley; pp. 28-33 and 310.

## XII. Annex.

##Here we will add Kirk's approximation

##and the Modified Kirk's approximation for spread option prices, but with a different K and rho

##Here we create a matrix for rho, with 21 rows, 1 column and 0.80, 0.85, 0.90, 0.95, 0.999 values

matrix_rho <- matrix(c(0.80, 0.85, 0.90, 0.95, 0.999), nrow = 5, ncol = 1)

##Here we create a function that will evaluate our formula

function_S3_K_rho<-function(rho){

##Here we simulate our Brownians

n = (5000000)

W = rnorm(n,mean=0,sd=1)

Z = rnorm(n,mean=0,sd=1)

##Here we find the correlation between W and Z

##rho = cor(W, Z)

##rho = 0.999

##Here we calculate B, so W and B can have same correlation

B = (rho * W + (sqrt(1-(rho^2))) * Z)

##Here we simulate our spreads

##Here we introduce the inputs for S1

##inputs:

r=0; sigma1=0.3; T=0.5; n=5000000; S0_1=100;

S1 = S0_1 * exp((r -(sigma1^2)/2)  * T + sigma1 * sqrt(T) * W)

##Here we introduce the inputs for S2

##inputs:

r=0; sigma2=0.2; T=0.5; n=5000000; S0_2=100;

S2 = S0_2 * exp((r -(sigma2^2)/2) * T + sigma2 * sqrt(T) * B)

## Here we invert our W and B values

invW = W * (-1)

invZ = Z * (-1)

invB = (rho * invW + (sqrt(1-(rho^2))) * invZ)

##Here we simulate our spreads

##Here we introduce the inputs for invS1

##inputs:

r=0; sigma1=0.3; T=0.5; n=5000000; S0_1=100;

invS1 = S0_1 * exp((r -(sigma1^2)/2)  * T + sigma1 * sqrt(T) * invW)

##Here we introduce the inputs for invS2

##inputs:

r=0; sigma2=0.2; T=0.5; n=5000000; S0_2=100;

invS2 = S0_2 * exp((r -(sigma2^2)/2) * T + sigma2 * sqrt(T) * invB)

##Here we create a matrix for K, with 21 rows, 1 column and values from 0 to 20

matrix_K <- matrix(c(0:20), nrow = 21, ncol = 1)

##Here we create a function that will evaluate our formula

function_S3_K <- function(K){

##Here we calculate max(S^1(T) - S^2(T) - 0, 0)

```r
S3 = pmax(S1 - S2 - K, 0)

S3_mean = mean(S3)

S3_mean

##Here we calculate max(S^1(T) - S^2(T) - 0, 0)

invS3 = pmax(invS1 - invS2 - K, 0)

invS3_mean = mean(invS3)

invS3_mean

S3_total_mean = (S3_mean + invS3_mean) / 2

S3_total_mean

}

##Here we apply the matrix_K_rho values to our function_total_S3 to be evaluated

results_function_S3_K = apply(matrix_K, 1, function(x) function_S3_K(x) )

}

results_function_S3_K_rho = apply(matrix_rho, 1, function(x) function_S3_K_rho(x) )

results_function_S3_K_rho

##Here we add Kirk's approximation for spread option prices

##Here we create a matrix for rho, with 21 rows, 1 column and 0.80, 0.85, 0.90, 0.95, 0.999 values

matrix_rho <- matrix(c(0.80, 0.85, 0.90, 0.95, 0.999), nrow = 5, ncol = 1)

##Here we create a function that will evaluate our formula

function_Kirk_K_rho<-function(rho){

##Here we create a matrix for K, with 21 rows, 1 column and values from 0 to 20

matrix_K <- matrix(c(0:20), nrow = 21, ncol = 1)
```

```r
##Here we create a function that will evaluate our formula

function_Kirk_K<-function(K){

##Here we calculate our volatility

##inputs:

r=0; sigma1=0.3; T=0.5; S0_1=100;

r=0; sigma2=0.2; T=0.5; S0_2=100;

a_kirk = sqrt( (sigma1)^2 - 2*rho*sigma1*sigma2 * (S0_2/(S0_2 + K)) + (sigma2)^2 * ((S0_2/(S0_2 + K))^2) )

##Here we calculate our d1 and d2

S = (S0_1 / (S0_2 + K))

d1 = (log(S, base = exp(1)) + 1/2 * (a_kirk^2) * T) / ( a_kirk * (sqrt(T)) )

d2 = d1 - a_kirk * sqrt(T)

##Here we use the above calculations to approximate the call spread using Kirk's formula

N_d1 = pnorm(d1,mean=0,sd=1)

N_d2 = pnorm(d2,mean=0,sd=1)

C_kirk = (exp(-r*T)) * ( (S0_1*N_d1) - ((S0_2 + K) * N_d2) )

C_kirk

}

##Here we apply the matrix_K_rho to our function_Kirk to be evaluated

results_function_Kirk_K = apply(matrix_K, 1, function(x) function_Kirk_K(x) )

}

results_function_Kirk_K_rho = apply(matrix_rho, 1, function(x) function_Kirk_K_rho(x) )

results_function_Kirk_K_rho
```

##Here we add Modified Kirk's approximation for spread option prices

##Here we create a matrix for rho, with 21 rows, 1 column and 0.80, 0.85, 0.90, 0.95, 0.999 values

matrix_rho <- matrix(c(0.80, 0.85, 0.90, 0.95, 0.999), nrow = 5, ncol = 1)

##Here we create a function that will evaluate our formula

function_Kirk_modif_K_rho<-function(rho){

##Here we create a matrix for K, with 21 rows, 1 column and values from 0 to 20

matrix_K <- matrix(c(0:20), nrow = 21, ncol = 1)

##Here we create a function that will evaluate our formula

function_Kirk_modif_K <- function(K){

##Here we calculate our volatility

##inputs:

r=0; sigma1=0.3; T=0.5; S0_1=100;

r=0; sigma2=0.2; T=0.5; S0_2=100;

a_kirk = sqrt( (sigma1)^2 - 2*rho*sigma1*sigma2 * (S0_2/(S0_2 + K)) + (sigma2)^2 * ((S0_2/(S0_2 + K))^2) )

X_t = log(S0_1, base = exp(1))

x_aster = log((S0_2 + K), base = exp(1))

I_t = sqrt(a_kirk^2) + 1/2 * (( (sigma2 * S0_2/(S0_2 + K)) - rho* sigma1)^2 ) * ( 1 / ((sqrt(a_kirk^2))^3) ) * (sigma2^2) * ( (S0_2 * K) / ((S0_2 + K)^2) ) * (X_t - x_aster)

S = (S0_1 / (S0_2 + K))

d1_modif = (log(S, base = exp(1)) + 1/2 * (I_t^2) * T) / (I_t * (sqrt(T)) )

d2_modif = d1_modif - I_t * sqrt(T)

N_d1_modif = pnorm(d1_modif,mean=0,sd=1)

N_d2_modif = pnorm(d2_modif,mean=0,sd=1)

```r
C_kirk_modif = (exp(-r*T)) * ( (S0_1*N_d1_modif) - ((S0_2 + K) * N_d2_modif) )

C_kirk_modif

}

##Here we apply the matrix_K_rho values to our function_Kirk_modif to be evaluated

results_Kirk_modif_K = apply(matrix_K, 1, function(x) function_Kirk_modif_K(x) )

}

results_function_Kirk_modif_K_rho = apply(matrix_rho, 1, function(x) function_Kirk_modif_K_rho(x) )

results_function_Kirk_modif_K_rho

results_function_S3_K_rho

results_function_Kirk_K_rho

results_function_Kirk_modif_K_rho

##Here we calculate the error produced by the Kirk formula for each rho

error_Kirk = ((results_function_Kirk_K_rho * 100) / results_function_S3_K_rho) - 100

colnames(error_Kirk) <- c('rho_0.8', 'rho_0.85','rho_0.9','rho_0.95', 'rho_0.999')

rownames(error_Kirk) <- c('K0', 'K1', 'K2','K3','K4','K5','K6','K7','K8','K9',
'K10','K11','K12','K13','K14','K15','K16','K17','K18','K19','K20')

error_Kirk

##Here we set and retrieve the current directory

setwd("~/")

getwd()

##Here we save the plot images

plot(error_Kirk[,'rho_0.8'], type= 'l', xlab="K", ylab='error_Kirk_0.8')

dev.print(pdf, 'error_Kirk_rho_0.8.pdf');
```

```r
plot(error_Kirk[,'rho_0.85'], type= 'l', xlab="K", ylab='error_Kirk_0.85')

dev.print(pdf, 'error_Kirk_rho_0.85.pdf');

plot(error_Kirk[,'rho_0.9'], type= 'l', xlab="K", ylab='error_Kirk_0.9')

dev.print(pdf, 'error_Kirk_rho_0.9.pdf');

plot(error_Kirk[,'rho_0.95'], type= 'l', xlab="K", ylab='error_Kirk_0.95')

dev.print(pdf, 'error_Kirk_rho_0.95.pdf');

plot(error_Kirk[,'rho_0.999'], type= 'l', xlab="K", ylab='error_Kirk_0.999')

dev.print(pdf, 'error_Kirk_rho_0.999.pdf');

dev.off ();

##Here we calculate the error produced by the modified Kirk formula for each rho

error_Kirk_modif = ((results_function_Kirk_modif_K_rho * 100) / results_function_S3_K_rho) - 100

colnames(error_Kirk_modif) <- c('rho_0.8', 'rho_0.85','rho_0.9','rho_0.95','rho_0.999')

rownames(error_Kirk_modif) <- c('K0', 'K1', 'K2','K3','K4','K5','K6','K7','K8','K9','K10','K11','K12','K13','K14','K15','K16','K17','K18','K19','K20')

error_Kirk_modif

##Here we save the plot images

plot(error_Kirk_modif[,'rho_0.8'], type= 'l', xlab="K", ylab='error_Kirk_modif_0.8')

dev.print(pdf, 'error_Kirk_modif_rho_0.8.pdf');

plot(error_Kirk_modif[,'rho_0.85'], type= 'l', xlab="K", ylab='error_Kirk_modif_0.85')

dev.print(pdf, 'error_Kirk_modif_rho_0.85.pdf');

plot(error_Kirk_modif[,'rho_0.9'], type= 'l', xlab="K", ylab='error_Kirk_modif_0.9')

dev.print(pdf, 'error_Kirk_modif_rho_0.9.pdf');

plot(error_Kirk_modif[,'rho_0.95'], type= 'l', xlab="K", ylab='error_Kirk_modif_0.95')
```

```r
dev.print(pdf, 'error_Kirk_modif_rho_0.95.pdf');

plot(error_Kirk_modif[,'rho_0.999'], type= 'l', xlab="K", ylab='error_Kirk_modif_0.999')

dev.print(pdf, 'error_Kirk_modif_rho_0.999.pdf');

dev.off ();

##Here we superpose the plot images and we save them

plot(error_Kirk[,'rho_0.8'],type="l",xlab="K",   ylab='error_Kirk_and_error_Kirk_modif_0.8', col="red", ylim=c(-0.1, 0.5))

lines(error_Kirk_modif[,'rho_0.8'],col="green", ylim=c(-0.5, 0.5))

legend('topright', legend=c("Kirk errors", "Modified Kirk errors"), col=c("red", "green"), lty=1, cex=0.8)

dev.print(pdf, 'error_Kirk_and_error_Kirk_modif_rho_0.8.pdf');

plot(error_Kirk[,'rho_0.85'],type="l",xlab="K",  ylab='error_Kirk_and_error_Kirk_modif_0.85', col="red", ylim=c(-0.2, 0.8))

lines(error_Kirk_modif[,'rho_0.85'],col="green", ylim=c(-0.2, 0.8))

legend('topright', legend=c("Kirk errors", "Modified Kirk errors"), col=c("red", "green"), lty=1, cex=0.8)

dev.print(pdf, 'error_Kirk_and_error_Kirk_modif_rho_0.85.pdf');

plot(error_Kirk[,'rho_0.9'],type="l",xlab="K",   ylab='error_Kirk_and_error_Kirk_modif_0.9', col="red", ylim=c(-0.2, 0.7))

lines(error_Kirk_modif[,'rho_0.9'],col="green", ylim=c(-0.2, 0.7))

legend('topright', legend=c("Kirk errors", "Modified Kirk errors"), col=c("red", "green"), lty=1, cex=0.8)

dev.print(pdf, 'error_Kirk_and_error_Kirk_modif_rho_0.9.pdf');

plot(error_Kirk[,'rho_0.95'],type="l",xlab="K",  ylab='error_Kirk_and_error_Kirk_modif_0.95', col="red", ylim=c(-0.4, 7))

lines(error_Kirk_modif[,'rho_0.95'],col="green", ylim=c(-0.4, 7))
```

legend('topright', legend=c("Kirk errors", "Modified Kirk errors"), col=c("red", "green"), lty=1, cex=0.8)

dev.print(pdf, 'error_Kirk_and_error_Kirk_modif_rho_0.95.pdf');

plot(error_Kirk[,'rho_0.999'],type="l",xlab="K", ylab='error_Kirk_and_error_Kirk_modif_0.999', col="red", ylim=c(-0.2, 5))

lines(error_Kirk_modif[,'rho_0.999'],col="green", ylim=c(-0.2, 5))

legend('topright', legend=c("Kirk errors", "Modified Kirk errors"), col=c("red", "green"), lty=1, cex=0.8)

dev.print(pdf, 'error_Kirk_and_error_Kirk_modif_rho_0.999.pdf');

dev.off ();